\documentclass[useAMS,usenatbib]{mn2e}

\usepackage{amssymb,amsmath}
\usepackage{graphicx}
\usepackage{color}
\bibpunct{(}{)}{;}{a}{}{,}
\usepackage{xspace}

\newcommand{\zacs}{$z_{850}$\xspace}
\newcommand{\hst}{HST\xspace}
\newcommand{\acs}{ACS\xspace}
\input{journals.dat}

\title[Brightest Cluster Galaxies]{The importance of major mergers in the build up of stellar mass in
 brightest cluster galaxies at $\mathbf{z=1}$}

\author[C. Lidman et al.]{C. Lidman$^{1}$\thanks{E-mail:
clidman@aao.gov.au}, G. Iacobuta$^{1,2}$, A. E. Bauer$^{1}$,
L. F. Barrientos$^{3}$, P. Cerulo$^{4}$, 
\newauthor W. J. Couch$^{4}$, L. Delaye$^{5}$, 
R. Demarco$^{6}$,  E. Ellingson$^{7}$A. J. Faloon$^{8}$, 
\newauthor D. Gilbank$^{9}$, M. Huertas-Company$^{5,10}$,
S. Mei$^{5,10}$, J. Meyers$^{11}$, A. Muzzin$^{12}$,
\newauthor  A. Noble,$^{8}$, J. Nantais$^{6}$, A. Rettura$^{13}$, P. Rosati$^{14}$,
R. S\'{a}nchez-Janssen$^{15}$, 
\newauthor V. Strazzullo$^{16}$, T. M. A. Webb$^{8}$, G. Wilson$^{17}$, R. Yan$^{18}$, H. K. C. Yee$^{19}$\\
$^{1}$Australian Astronomical Observatory, PO Box 915, North Ryde, NSW 1670, Australia\\
$^{2}$School of Physics and Astronomy, University of Nottingham, University Park, Nottingham, NG7 2RD, UK\\
$^{3}$Departamento de Astronom\'{i}a y Astrof\'{i}sica Pontifica Universidad Cath\'{o}lica de Chile, Vicu\~{n}a MacKenna 4860, \\
$^{4}$Centre for Astrophysics and Supercomputing, Swinburne University of Technology, PO Box 218 Hawthorn, VIC 3122 Australia\\
$^{5}$GEPI, Observatoire de Paris, 77, Avenue Denfert-Rochereau, Paris, France\\
$^{6}$Department of Astronomy, Universidad de Concepcion, Casilla 160-C Concepcion, Chile\\
$^{7}$Center for Astrophysics and Space Astronomy, Department of Astrophysical and Planetary Science, UCB-389, \\
\hspace{0.27cm}University of Colorado, Boulder, CO, 80309, USA\\
$^{8}$McGill University, 3600 rue University, Montreal, QC, Canada, H3A 2T8\\
$^{9}$South African Astronomical Observatory, P.O. Box 9, Observatory, 7935, South Africa\\
\hspace{0.27cm}7820436 Macul, Santiago, Chile\\
$^{10}$Universit\'{e} Paris Denis Diderot, 75205 Paris Cedex13, France\\
$^{11}$Department of Physics, Stanford University, 450 Serra Mall  Stanford, CA 94305, USA\\
$^{12}$Leiden Observatory, Leiden University,, PO Box 9513, 2300 RA Leiden, The Netherlands\\
$^{13}$Jet Propulsion Laboratory, 4800 Oak Grove Dr  Pasadena, CA 91109, USA\\
$^{14}$European Southern Observatory, Karl-Schwarzschild Strasse 2, 85748 Garching, Germany\\
$^{15}$European Southern Observatory, Alonso de Cordova 3107, Vitacura, Santiago, Chile\\
$^{16}$Laboratoire AIM-Paris-Saclay, CEA/DSM-CNRS, Université Paris
Diderot, Irfu/Service d'Astrophysique, \\ \hspace{0.27cm}CEA Saclay, Orme des Merisiers,
F-91191 Gif sur Yvette, France\\
$^{17}$Department of Physics and Astronomy, University of California, Riverside, CA 92521, USA\\
$^{18}$Department of Physics and Astronomy, University of Kentucky, 505 Rose St. Lexington, KY, 40506-0055, USA\\
$^{19}$Department of Astronomy and Physics, University of Toronto,
Toronto, Ontario, M5S 3H4, Canada\\
}

\begin{document}

\date{Accepted 2013 May 1. Received 2013 April 18; in original form 2013 March 7}

\pagerange{\pageref{firstpage}--\pageref{lastpage}} \pubyear{2002}

\maketitle

\label{firstpage}

\begin{abstract}
  Recent independent results from numerical simulations and
  observations have shown that brightest cluster galaxies (BCGs) have
  increased their stellar mass by a factor of almost two between
  $z\sim0.9$ and $z\sim0.2$. The numerical simulations further suggest
  that more than half this mass is accreted through major
  mergers. Using a sample of 18 distant galaxy clusters with over 600
  spectroscopically confirmed cluster members between them, we search
  for observational evidence that major mergers do play a significant
  role. We find a major merger rate of $0.38\pm0.14$ mergers per Gyr
  at $z \sim 1$.  While the uncertainties, which stem from the small
  size of our sample, are relatively large, our rate is consistent
  with the results that are derived from numerical simulations.  If we
  assume that this rate continues to the present day and that half of
  the mass of the companion is accreted onto the BCG during these
  mergers, then we find that this rate can explain the growth in the
  stellar mass of the BCGs that is observed and predicted by
  simulations. Major mergers therefore appear to be playing an
  important role, perhaps even the dominant one, in the build up of
  stellar mass in these extraordinary galaxies.

\end{abstract}

\begin{keywords}
galaxies: clusters: general, galaxies: evolution, galaxies: high-redshift
\end{keywords}

\section{Introduction}

Brightest cluster galaxies (BCGs) are amongst the largest, most
massive and most luminous galaxies in the universe.  Often found close
to the centre of the cluster that they inhabit, BCGs are generally
easy to identify, even in the most distant clusters that are currently
known. They are also easy to identify in large N-body simulations,
thus allowing us the opportunity to directly compare, for a single
class of galaxy, the predictions of numerical simulations with
observations.

In the hierarchical scenario for the formation of structure in our
universe, BCGs build up their stellar mass over time by converting
material accreted from their surroundings into stars and by merging
with other galaxies. Over the range of redshifts that we can observe
BCGs, the stellar mass of the average BCG is expected to increase
significantly with time through merging with other galaxies. For
example, in the semi-analytic model described in \citet{DeLucia2007},
the stellar mass increases by a factor of four between redshift
$z=1.0$ and today.

Observationally, it appears that the growth is slightly slower than
the predictions of the \citet{DeLucia2007} model. From a sample of 150
BCGs, \citet{Lidman2012} found that the stellar mass of BCGs increases
by a factor 1.8 over the redshift interval $z\sim 0.9$ to
$z\sim0.2$. This result differs from that of earlier works, which
generally found little or no change over the same redshift interval
\citep{Whiley2008,Stott2008,Collins2009,Stott2010}.  In part, this was
due to the way the positive correlation between the mass of the
cluster and the stellar mass of the BCG
\citep{Edge1991,Burke2000,Brough2002,Brough2008,Stott2008} tends to
dilute the observed evolution. The distant clusters used in these
samples were more massive than the likely progenitors of clusters in
the low-redshift comparison samples. The distant clusters therefore
tended to have more massive BCGs to start with. 

More recent models \citep{Laporte2013} predict that BCGs
should increase their stellar mass by a factor of 1.9 between $z=1.0$ and $z=0.2$, an
increase that is lower than that reported in earlier simulations \citep{DeLucia2007}. The
theoretical expectation and the observations are now in excellent agreement.
The new models also predict that the growth occurs through a combination of
minor and major mergers\footnote{Throughout this paper we use
  $0.25 <\mu_{\star} <1$ to define a  major merger, where $\mu_{\star}
  $ is the mass ratio between the satellite and its more
  massive companion.} and that the size of BCGs should increase dramatically as they
grow in mass.


Observational support for the notion that BCGs build up their stellar
mass by merging has been steadily increasing over the past few
years. From a sample of 91 BCGs at $z\sim 0.3$, \citet{Edwards2012}
found that BCGs increase their mass by as much as 10\% over 0.5\,Gyr.
Both minor and major mergers  are thought to play a role, with major mergers
contributing somewhere between half \citep{Edwards2012} to
substantially more than half \citep{Hopkins2010a,Laporte2013} of the mass. Direct
evidence for merging, through tidal tails and distorted isophotes, has
been found by a number of authors
\citep{McIntosh2008,Rasmussen2010,Liu2009,Brough2011,Bildfell2012}. \citet{Liu2009}
estimated that 3.5\% of BCGs in the redshift interval $0.03 \le z \le
0.12$ show evidence for an ongoing merger. \citet{McIntosh2008} found a
similar fraction and estimated that the centres of groups and clusters
are increasing their mass at a rate of 2-9\% per Gyr.
Once signs of a merger are evident, the timescale for merging is
short, of the order of a few crossing times, which is typically around
0.2\,Gyr for galaxies that are within 30\,kpc of the BCG.

The amount of mass accreted through mergers by BCGs in distant
clusters, i.e.~those at $z\sim 1$, is largely unconstrained. 
A measurement of the mass accreted by high redshift BCGs can
be combined with the measurements at low redshift, and allow us to
estimate the mass accreted through mergers from redshift 1 to today.

In this paper, we combine high-quality ground-based near-IR images of
a sample of 18 distant galaxy clusters with extensive spectroscopy to
examine the possibility that major mergers between the brightest
galaxies within the core of the cluster and the BCG contribute
significantly to the growth of the stellar mass of the BCG between $z
\sim 1$ and today. In Sec.~\ref{sec:data}, we introduce the sample
that we use in the analysis, which includes new high-resolution data
taken with the HAWK-I\footnote{High Acuity Wide field K-band Imager}
camera on the Very Large Telescope (VLT). Since much of the HAWK-I
data have not been published before and since we will make these data
public, we provide a detailed description of how the these data were
taken and processed. After expanding the work presented in
\citet{Lidman2012} by adding new measurements to the rest-frame K-band
magnitude-redshift relation in Sec.~\ref{sec:Hubble}, we use the
radial distribution of over 600 spectroscopically confirmed cluster
galaxies, both bright and faint, to argue in Sec.~\ref{sec:companions}
that there is an excess of bright galaxies close to the BCG.  We then
estimate the timescale for the brighter galaxies to merge with the BCG
and then infer how many close companions we should have seen if major
merging is the dominant process for the build-up of stellar mass in
BCGs.  In the final two sections of the paper, we discuss and
summarise our main results.

Throughout the paper, all magnitudes and colours are measured in the
observer frame and are placed on the 2MASS photometric system. Vega
magnitudes are used throughout. We also assume a flat cold dark matter
cosmology with $\Omega_{\Lambda}=0.73$ and $H_0=70$
km\,s$^{-1}$\,Mpc$^{-1}$. 

\section{The Cluster Sample} \label{sec:data}
  
The sample of clusters we use in this paper is built from several surveys. Ten clusters
come from SpARCS\footnote{Spitzer Adaptation of the Red-Sequence
  Cluster Survey,
  www.faculty.ucr.edu/$\sim$gillianw/SpARCS/}\citep{Muzzin2009a,
  Wilson2009a}. In brief, the SpARCS clusters were discovered by
searching for over-densities in the number of red galaxies using
images taken with IRAC on the Spitzer Space Telescope and ground-based
z-band images taken with either MegaCam on the Canada-France-Hawaii
Telescope (CFHT) or MOSAIC II on the Cerro Tololo Blanco
Telescope. Additional details on individual clusters can be found in
\citet{Muzzin2009a}, \citet{Wilson2009a}, and \citet{Demarco2010a}.

All ten clusters from SpARCS were also part of GCLASS\footnote{Gemini
  Cluster Astrophysics Spectroscopic Survey,
  www.faculty.ucr.edu/$\sim$gillianw/GCLASS/}, a spectroscopic survey
that used the multi-object spectroscopic (MOS) modes of GMOS-N and
GMOS-S on the Gemini Telescopes to obtain between 20 and 80
spectroscopically confirmed members per cluster
\citep{Muzzin2012}. The comprehensive spectroscopic coverage provided
by GCLASS allows us to identify the brightest cluster members in each
of the clusters, and to exclude foreground and background galaxies
that might be confused as cluster members. It is for this reason that
we do not add the two SpARCS clusters at $z \sim 1.63$ in
\citet{Lidman2012} to the sample. The number of spectroscopically
confirmed cluster members is currently around a dozen for both clusters,
although work to increase this number significantly is currently
underway.

Complementing the comprehensive spectroscopic coverage are
ground-based images in the optical (u, g, r, i and z) and near-IR (J and
Ks), images in each of the IRAC passbands ([3.6], [4.5], [5.8] and
[8]), and for some clusters, images with MIPS.  The
near-IR data are described in \citet{Lidman2012}, while
the optical data are described in van der Berg et al. (in preparation).

The remaining clusters were discovered either through their X-ray
emission or as over-densities of red galaxies. Most of these clusters
were observed with HAWK-I as part of the HAWK-I cluster survey, which
we describe in greater detail in the following section. All clusters
are listed in Table~\ref{tab:clustersample}. The redshift range
covered by the sample extends from z=0.84 to z=1.46.

\subsection{The HAWK-I cluster survey}\label{sec:HCS}

During 2005 and 2006, the Supernova Cosmology Project (SCP) targeted 25
galaxy clusters in the redshift range $0.9 < z < 1.5$ with the \acs
camera on \hst with the purpose of finding distant Type Ia supernovae
(SNe Ia). The survey was called the HST Cluster Supernova Survey, and
is described in \citet{Dawson2009}.  Given the relatively small
fields-of-view that are available with HST and the high surface
density of potential hosts in distant clusters, distant clusters are
an efficient way of finding distant SNe Ia \citep{Dawson2009,Suzuki2012}.

Clusters for the HST Cluster Supernova Survey were selected from the
IRAC Shallow Cluster Survey \citep{Eisenhardt2008}, the Red-sequence
Cluster Surveys \citep[RCS and RCS-2,][]{Gladders2005,Yee2007}, the
XMM Cluster Survey \citep{Sahlen2009}, the Palomar Distant Cluster
Survey \citep{Postman1996}, the XMM-Newton Distant Cluster Project
\citep{Boehringer2005}, and the ROSAT Deep Cluster Survey
\citep[RDCS,][]{Rosati1999}. At the time the HST Cluster Supernova Survey was
conducted, the sample represented a significant fraction of the known
$z>0.9$ clusters.
 
The spectroscopic follow-up of supernovae in these clusters was done with
FORS2 on the VLT, LRIS on Keck and FOCAS on Subaru. Details on the
follow-up can be found in \citet{Dawson2009} and
\citet{Morokuma2010}. In most cases, the MOS mode was used, which
allowed one to obtain the redshifts of many cluster members in
addition to that of the host of the supernova or the supernova
itself. Some clusters, such as XMMU~1229 \citep{Santos2009}, produced several SNe, which
meant that they were targeted multiple times, thereby leading to many
redshifts. 

Many of the clusters in the HST Cluster Supernova Survey have been 
observed at other wavelengths. This includes longer wavelength data 
taken with IRAC and MIPS on the Spitzer Space Telescope and, for a few 
clusters, data from the Herschel Space Observatory.  This broad coverage, 
together with the large number of spectroscopically confirmed cluster 
members in each cluster means that these clusters are an ideal sample 
for studying the properties of galaxies in distant clusters. Lacking, 
however, were high-quality near-IR data that matched the depth of the 
data taken in space. This was the driving reason for observing many of
these clusters with HAWK-I on the VLT in a survey that we refer to in
the rest of the paper as the HAWK-I cluster survey (HCS).

Clusters for the HCS were selected from those targeted in the HST Cluster
Supernova Survey according to two criteria:

\begin{itemize}

\item They were visible from the Paranal Observatory

\item There were at least 10 spectroscopically confirmed members per cluster

\end{itemize}

To this list, we added RX~J0152.7-1357 (hereafter, RX~0152) a
X-ray discovered cluster at z=0.84 that has over 100
spectroscopically confirmed cluster members \citep{Demarco2010b} and
deep ACS imaging \citep{Blakeslee2006}. 

Six of the nine clusters were discovered from their X-ray
emission. One of these, XLSS~0223, was independently discovered as an
overdensity of galaxies in images taken with IRAC \citep{Bremer2006, Muzzin2013}. The
remainder were discovered as over-densities of red-sequence galaxies in
the Red-sequence Cluster Survey \citep[RCS,][]{Gladders2005}. 


The complete list of 19 clusters, which includes the
clusters from SpARCS, the method by which they were
discovered, their redshifts, and a list of selected references
are listed in Table~\ref{tab:clustersample}. Also listed are the
abbreviated names that we use throughout the text.


 \begin{table*}
 \centering
  \caption{The 19 clusters in our initial sample. Listed first are clusters from
    the HAWK-I cluster survey, followed by clusters from SpARCS \label{tab:clustersample}}
  \begin{tabular}{llrlrl}
  \hline
   Name     &  Abbreviated name & Redshift           & Discovery
   method  &  Members$^{\mathrm {a}}$ & References\\
              
\hline
               RX~J0152.7-1357 & RX~0152 & 0.8360 & X-ray & 109 & \citet{Demarco2005}\\
             RCS~231953+0038.0 & RCS~2319 & 0.9024 & Optical & 25 & \citet{Gilbank2008}\\
             XMMU~J1229.4+0151 & XMMU~1229 & 0.9755 & X-ray & 18 & \citet{Santos2009}\\
             RCS~022056-0333.4 & RCS~0220 & 1.0271 & Optical & 7 & \citet{Munoz2009}\\
             RCS~234526-3632.6 & RCS~2345 & 1.0360 & Optical & 29 & \citet{Munoz2009}\\
             XLSS~J0223.0-0436 & XLSS~0233 & 1.2132 & X-ray/Optical-IR & 20 & \citet{Bremer2006}\\
             RDCS~J1252.9-2927 & RDCS~1252 & 1.2380 & X-ray & 42 & \citet{Rosati2004}\\
             XMMU~J2235.3-2557 & XMMU~2235 & 1.3900 & X-ray & 25 & \citet{Mullis2005}\\
           XMMXCS~J2215.9-1738 & XMMXCS~2215 & 1.4600 & X-ray & 26 & \citet{Stanford2006}\\
 \hline
SpARCS J003442-430752 & SpARCS~0034 &0.867 & Optical-IR & 39 & \citet{Muzzin2012}\\
SpARCS J003645-441050 & SpARCS~0036 &0.869 & Optical-IR & 46 & \citet{Muzzin2012}\\
SpARCS J161314+564930 & SpARCS~1613 &0.871 & Optical-IR & 87 &\citet{Demarco2010a} \\
SpARCS J104737+574137 & SpARCS~1047 &0.956  & Optical-IR& 26 &\citet{Muzzin2012}\\
SpARCS J021524-034331 & SpARCS~0215 &1.004  & Optical-IR & 42 &\citet{Muzzin2012}\\
SpARCS J105111+581803 & SpARCS~1051 &1.035  & Optical-IR & 26 & \citet{Muzzin2012}\\
SpARCS J161641+554513 & SpARCS~1616 &1.156  & Optical-IR & 37 &\citet{Demarco2010a} \\
SpARCS J163435+402151 & SpARCS~1634 &1.177  & Optical-IR & 35 & \citet{Muzzin2009a}\\
SpARCS J163852+403843 & SpARCS~1638 &1.195  & Optical-IR & 18 &\citet{Muzzin2009a}\\
SpARCS J003550-431224 & SpARCS~0035 &1.335  & Optical-IR & 21& \citet{Wilson2009a}\\
\hline
\end{tabular}
\begin{list}{}{}

\item[$^{\mathrm{a}}$]  The number of spectroscopically
    confirmed members within $r_{200}$\ of the cluster centre and with
    peculiar velocities that are less than three times the cluster
    velocity dispersion. $r_{200}$ is the radius within which the mean
    density of the cluster equals the critical density of the Universe
    at the redshift of the cluster multiplied by a factor of 200.

\end{list}

\end{table*}

\subsection{Observations with HAWK-I}

Eight of the nine clusters in the HCS were imaged with HAWK-I on Yepun
(VLT-UT4) at the ESO Cerro Paranal Observatory. HAWK-I
\citep{Pirard2004a,Casali2006a} is a near-IR imager with a 7\farcm5 x
7\farcm5 field of view. The focal plane consists of a mosaic of 4
Hawaii-2RG detectors and results in an average pixel scale of
0\farcs1065 per pixel. All clusters were imaged in Ks. Clusters above
redshift $z=1.2$ were also imaged in J in order to have two filters,
in tandem with \zacs, closely bracketing the 4000$\AA$ break. At
$z\sim 1.2$, the 4000$\AA$ break starts to move out of the ACS \zacs
bandpass.

One other cluster, RDCS~1252\footnote{We use abbreviated names
  throughout the paper. The full names are given in
  Table~\ref{tab:clustersample}}, had existing deep ISAAC J and Ks
band data. The ISAAC data on RDCS~1252 are exceptionally deep
\citep{Lidman2004}, so additional data with HAWK-I were not needed.

In the remainder of this section we refer to the seven clusters that
were observed as part of ESO program 084.A-0214. This excludes RDCS~1252
and XMMU~2235.  The near-IR data on these clusters are fully described
in \citet{Lidman2004} and \citet{Lidman2008}.

In order to cover a wide area and to keep the clusters away from the
gaps between the detectors, the observations were not done with the
cluster positioned in the centre of the mosaic. Instead, a pair
of pointings with the cluster roughly centred in quadrants 1 and 3 of the
mosaic was used. The two quadrants were chosen as they have the
highest quantum efficiency. The resulting union of images covers 10\arcmin\ by
10\arcmin\ of the sky.

Individual exposures lasted 20 seconds in J and 10 seconds Ks, and 6
of these were averaged to form a single image. Between images, the
telescope was moved by 10\arcsec\ to 30\arcsec\ in a semi-random
manner, and 23 (40 for Ks) images were taken in this way in a single
observing block (OB). The sequence was repeated several times.  Total
exposure times, detection limits and other observing details are
reported in Table \ref{tab:observations}.

Zero points were set using stars from the 2MASS point source catalogue
\citep{Skrutskie2006}. Typically, around 10-20 unsaturated 2MASS
stars with 2MASS quality flags of 'A' or 'B' were selected to measure
zero points and their uncertainties.  2MASS stars were weighted by the
reported uncertainties in the 2MASS point source catalogue. Standard
stars from \citet{Persson1998} were observed on some of the nights our
fields were observed. The agreement between the zero points derived
from the standards and the zero points derived using 2MASS stars was
always better than 5\% and was generally around 2\%.

\subsection{Data processing}

The processing of the raw data was done in a standard manner and
largely follows the steps outlined in \citet{Lidman2008}. A few minor
refinements were made. As in \citet{Lidman2008}, SCAMP (version 1.6.2)
and SWARP (version 2.17.6)\footnote{http://www.astromatic.net/} were
used to place the images onto a common astrometric reference frame;
however, we did not use Swarp to combine images. Instead, we used the
IRAF\footnote{IRAF is distributed by the National Optical Astronomy
  Observatories which are operated by the Association of Universities
  for Research in Astronomy, Inc., under the cooperative agreement
  with the National Science Foundation} task imcombine to combine the
images processed by SWarp. We weighted the images that went into the
combined image with the inverse square of the FWHM of the PSF.

Both the HAWK-I data on XMMU~2235, presented in \citet{Lidman2008}, and
the ISAAC data on RDCS~1252, presented in \citet{Lidman2004}, were
reprocessed to match the processing done here.

The quality of data, as measured by the image quality in the stacked
images and the depth of the images is very high. The image quality
(FWHM) is never poorer than 0\farcs5 and is often better than
0\farcs4, while the 5\,$\sigma$ detection limits in the Ks-band are
5--6 mag fainter than the BCGs.


 \begin{table*}
 \centering
 \begin{minipage}{140mm}
  \caption{Observing Log\label{tab:observations}}
  \begin{tabular}{lrlrll}
  \hline
  Name     & Instrument & Filter  & Exposure time & Image quality & 5
  $\sigma$ detection limit$^{\mathrm {a,b}}$\\
           &   & & [s] & [\arcsec] & [mag]\\
\hline
               RX~J0152.7-1357 &   HAWK-I & Ks &  9600 & 0.35 & 22.8 \\
             RCS~231953+0038.0 &   HAWK-I & Ks &  9600 & 0.40 & 22.6 \\
             XMMU~J1229.4+0151 &   HAWK-I & Ks & 11310 & 0.35 & 23.2 \\
             RCS~022056-0333.4 &   HAWK-I & Ks &  9600 & 0.31 & 23.1 \\
             RCS~234526-3632.6 &   HAWK-I & Ks &  9600 & 0.35 & 22.9 \\
             XLSS~J0223.0-0436 &   HAWK-I &  J & 11040 & 0.32 & 24.5 \\
             XLSS~J0223.0-0436 &   HAWK-I & Ks &  9600 & 0.32 & 22.9 \\
             RDCS~J1252.9-2927 &    ISAAC & Js & 86640 & 0.44 & 25.1 \\
             RDCS~J1252.9-2927 &    ISAAC & Ks & 81990 & 0.38 & 23.5 \\
             XMMU~J2235.3-2557 &   HAWK-I &  J & 10560 & 0.46 & 24.0 \\
             XMMU~J2235.3-2557 &   HAWK-I & Ks & 10740 & 0.31 & 22.5 \\
           XMMXCS~J2215.9-1738 &   HAWK-I &  J & 14400 & 0.47 & 24.1 \\
           XMMXCS~J2215.9-1738 &   HAWK-I & Ks &  9600 & 0.35 & 23.0 \\
\hline
\end{tabular}
\begin{list}{}{}

\item[$^{\mathrm{a}}$] All quantities refer to the central part of
  each image, where the image depth is greatest.

\item[$^{\mathrm{b}}$] The detection limit is the 5\,$\sigma$
  point-source detection limit within an aperture that has a diameter
  equal to twice the image quality. It takes into account the
  correlation in the noise between pixels and does not include the
  flux that falls outside the aperture.
\end{list}
\end{minipage}
\end{table*}

\subsection{Spectroscopic completeness}\label{sec:ID}

We noted earlier that there has been extensive spectroscopic follow-up
of the clusters in our sample, leading to a large number of objects
with redshifts. To estimate the redshift completeness, we first
construct colour-magnitude diagrams of all clusters for which we have
J and Ks images of comparable quality and depth (13 clusters in
total), and then select all galaxies that are within 0.2\,mag of the
red-sequence. We take this galaxy subsample and count the number of
galaxies that have redshifts (whether they are cluster members or not)
and the number of galaxies that do not have redshifts. We compute
these numbers in five bins.  The boundaries of the bins correspond to
six flux ratios: galaxies that are three times brighter (as measured in
the observer--frame Ks band) than the BCG, galaxies that are as bright
as the BCG and galaxies that are one-half, one-third, one-quarter and
one-tenth as bright as the BCG. We then compute the fraction of
galaxies that have redshifts (the spectroscopic completeness) for each
of these bins.

The completeness depends on the radius within which one chooses to
count galaxies.  It reaches an average of about 90\% at a 250\,kpc for
objects that brighter than one-quarter of the brightness of the BCG,
and then steadily drops with increasing radius. To demonstrate the
completeness, we choose two radii: 250\,kpc and 500\,kpc. The results
are displayed in Table~\ref{tab:completeness}.

For galaxies that are brighter than the BCG, we are (85) 100\% complete
within (500) 250\,kpc of the cluster centre. The completeness steadily drops
as one goes to fainter magnitudes. In the following section, we
discuss how the spectroscopic incompleteness may bias the choice of
which galaxy is the BCG.

In addition to examining the entire sample, we examined the SpARCS and
HCS samples separately, since the spectroscopic follow-up of the HCS
and SpARCS clusters differ. The spectroscopic follow-up of the SpARCS
clusters involved a single program (GCLASS) using GMOS-South and
GMOS-North at the Gemini Observatory \citep{Muzzin2012}; whereas the
follow-up of the HCS clusters involved multiple instruments at
multiple observatories and was spread over about 10 years. Broadly
speaking, the completeness of the two samples are similar, with GCLASS
being  slightly more complete in the three faintest bins.

\begin{table}
 \centering
  \caption{Percentage of galaxies near to the red-sequence  and within 250\,kpc and
    500\,kpc of the
    cluster centre with redshifts.\label{tab:completeness}}
  \begin{tabular}{ccc}
  \hline
 Brightness range$^{\mathrm{a}}$            & \multicolumn{2}{c}{Completeness}\\
                   & (250\,kpc)  & (500\,kpc)  \\
\hline
1--3           & 100  & 85\\
0.5--1        & 93    & 90 \\
0.33--0.5   & 81    & 71 \\
0.25--0.33 & 80    & 76 \\
0.1--0.25   & 67    & 64 \\
\hline
0.25--1    & 85 & 78 \\
\hline
\end{tabular}
\begin{list}{}{}

\item[$^{\mathrm{a}}$] The range is relative to the luminosity of
  the BCG in the Ks band. For example, 1--3  refers to galaxies that are between one to three
  times brighter than the BCG.

\end{list}

\end{table}

\section{Brightest Cluster Galaxies}\label{sec:Hubble}

\subsection{Selecting the Brightest Cluster Galaxy}\label{sec:ID}

From our initial sample of 19 clusters, we exclude RCS~0220.  The
central region of RCS~0220 is partially obscured by a nearby, almost
face-on spiral galaxy. The spiral galaxy adds an extra degree of
uncertainty in identifying the BCG and measuring its flux, because the
BCG could be obscured by the spiral. We note that none of our
conclusions change if we had chosen to keep this cluster.

From the 18 clusters that remain, we then use the following criteria
to select the brightest cluster galaxy.

\begin{itemize}

\item The velocity of the galaxy relative to the systemic velocity of
  the cluster is less than three times the cluster velocity
  dispersion.

\item The galaxy lies within $r_{200}$ of the luminosity weighted
  centroid of spectroscopically confirmed cluster members. We compute
  $r_{200}$ using either X-ray derived masses, if available, or the measured velocity
  dispersion, if not.

\item The galaxy is the brightest galaxy in the Ks band that remains.

\end{itemize}

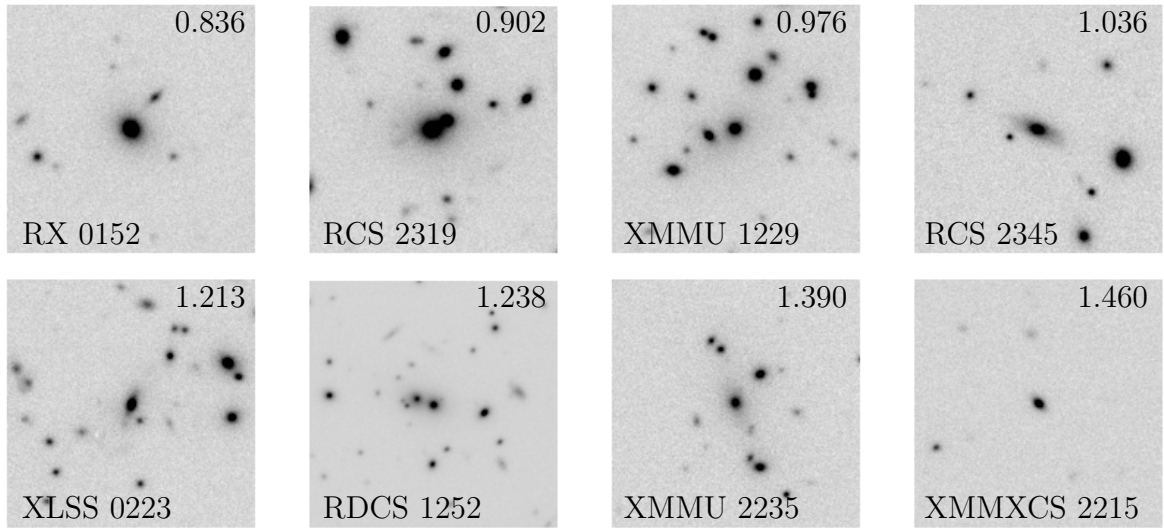
\begin{figure*}
\setlength{\unitlength}{0.3125mm}
\begin{picture}(512,256)
\includegraphics[width=16cm]{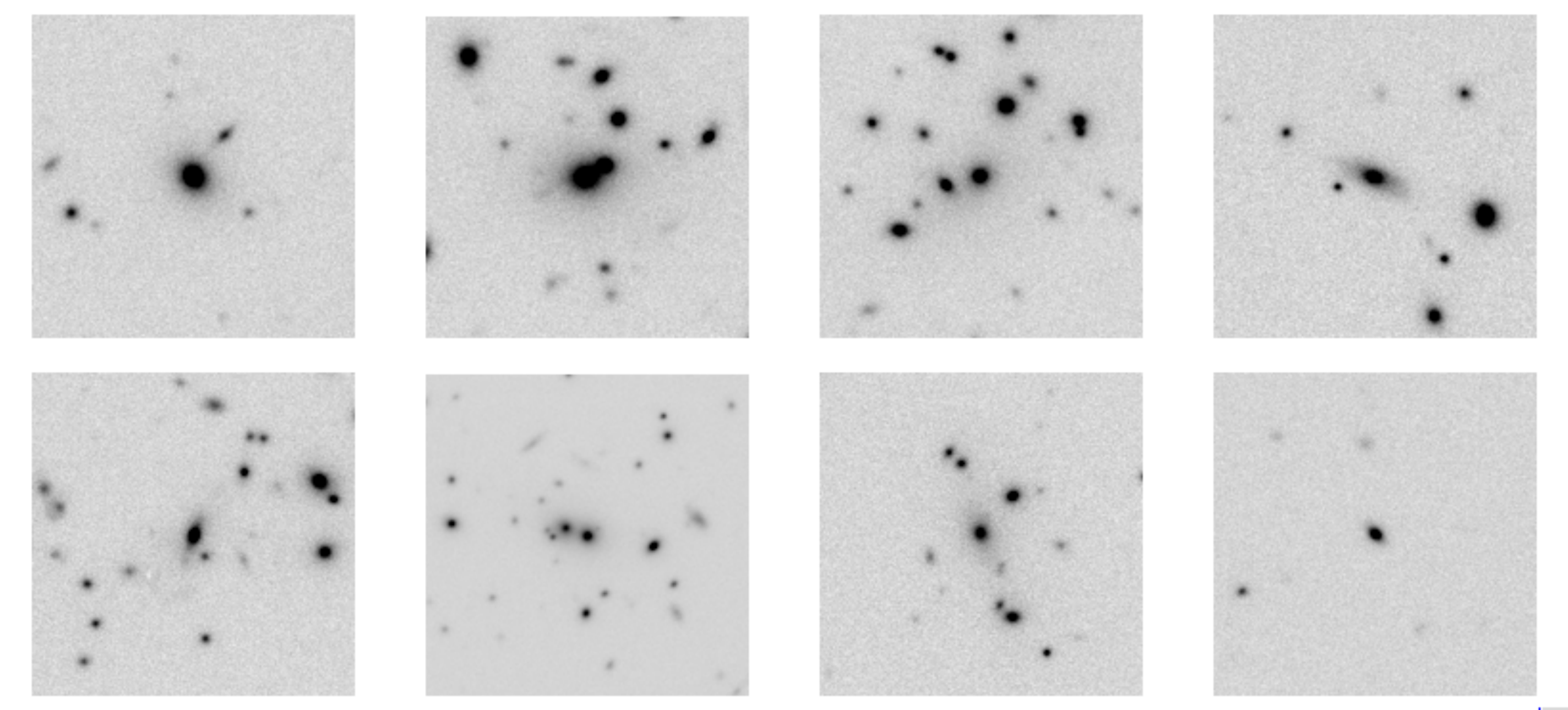}
\put(-430,215){\Large{0.836}}
\put(-302,215){\Large{0.902}}
\put(-174,215){\Large{0.976}}
\put(-46,215){\Large{1.036}}
\put(-430,98){\Large{1.213}}
\put(-302,98){\Large{1.238}}
\put(-174,98){\Large{1.390}}
\put(-46,98){\Large{1.460}}
\put(-495,125){\Large{RX~0152}}
\put(-367,125){\Large{RCS~2319}}
\put(-239,125){\Large{XMMU~1229}}
\put(-111,125){\Large{RCS~2345}}
\put(-495,8){\Large{XLSS~0223}}
\put(-367,8){\Large{RDCS~1252}}
\put(-239,8){\Large{XMMU~2235}}
\put(-111,8){\Large{XMMXCS~2215}}
\end{picture}
\caption{ Ks-band image cutouts centred on the HCS BCGs. With the
  exception of RDCS~1252, the images are 18\arcsec\ on a side, which
  corresponds to 140\,kpc for the nearest BCG and 155\,kpc for the
  most distant. For RDCS~1252, the image covers twice the area.
  Annotating each image are the shortened version of the cluster name
  and the redshift of the cluster. From left to right and from top to
  bottom, clusters are ordered in redshift. North is up and east is to
  the left. }\label{fig:BCGimages}
\end{figure*}

The BCGs of the SpARCS clusters are discussed in \citet{Lidman2012},
so here we concentrate on the BCGs in the HCS clusters. The BCGs in the HCS
generally within 100\,kpc of the centres of their respective
clusters. There are few notable exceptions, which we discuss
here. Thumbnails centered on the BCGs in the HCS clusters are shown in
Fig.~\ref{fig:BCGimages} and coordinates of the BCGs are listed in
Table~\ref{tab:BCGphot}.


RX~0152 consists of at least three dynamically distinct clumps
\citep{Demarco2005, Demarco2010b}. The two main clumps, which we refer
to as the northern and southern clumps, are separated from each other by
about 1\farcm5 (700\,kpc). Both clumps emit in the X-ray
\citep{Demarco2005}.  For this cluster, the BCG is not centered in
either of the clumps but centered about 35\arcsec\ (270\,kpc)
north-east of the northern clump. It is interesting to note that the
brightest galaxies that are centered in these clumps share a common
characteristic. Neither one is the BCG of the cluster and both have
bright nearby companions. The companion of the galaxy that is centered
in the southern clump is only 0\farcs5 (3.5\,kpc) away. These two
galaxies were selected as the BCG in \citet{Stott2010}. We discuss
RX~0152 further in Sec.~\ref{sec:discussion}.


XLSS~0223 contains two galaxies in the core of the cluster that differ
by only 0.02 mag. The projected separation of the two galaxies is
$\sim 60$\,kpc. The brighter of the two, which is selected as the BCG in
this study, is considerably more disk-like. In the higher-resolution
ACS images, the isophotes are distorted, which is perhaps an
indication of an ongoing merger. 

XMMXCS~2215 contains several galaxies within the core that are almost
as bright as one another. The galaxy we selected as the BCG is about
15\arcsec north-west of the cluster centre and is the same galaxy
selected in earlier works \citep{Collins2009,Stott2010}. The choice
is not unambiguous, as there are several galaxies in this cluster with
a similar magnitude. An observation in a different filter may have
resulted in a different choice.  XMMLSS~2215 is also interesting for
another reason. About 2\farcm3 south of the cluster there is a galaxy
that is even brighter than the galaxy that we have selected as the
BCG. It is not selected as the BCG of the cluster as it lands outside
$r_{200}$.  It is close to a tight knot of galaxies that currently
lack redshifts, so we do not know if this knot of galaxies are
associated to the main cluster.

RCS~2345 shares a few similarities to XMMXCS~2215. The cluster is
relatively open and there are several cluster members throughout the core
and beyond with similar magnitudes.  The galaxy that fulfils our
definition, as the BCG is about 1\arcmin\ (500\,kpc) to the south of
what appears to be the core of the cluster. It is the most isolated
BCG in our sample.

We note that modifying our BCG selection criteria by
reducing the radius over which BCGs are selected -- from $r_{200}$ to
$r_{500}$ -- and by treating the two clumps of RX0152 as separate
clusters would not affect our sample of BCGs significantly. Apart from
gaining an additional BCG from the southern clump in RX~0152, only the
BCG in RCS~2345 would change.

It is worth reflecting on how robust our selection is to spectroscopic
incompleteness. If the BCG occurs within 250\,kpc of the cluster
center and has a J-Ks colour that places it within 0.2\,mag of the red
sequence, then we will have selected it, as our spectroscopy is 100\%
complete for objects that are up to three times brighter than the
object that was chosen as the BCG. If it lies beyond this radius, then
there is a small chance that we would not have selected it.  Within
500\,kpc of the cluster centres, our spectroscopy for objects that are
up to three times as bright as the chosen BCG is 85\% complete. As
noted earlier, only the BCG of RCS~2345 is significantly more than
250\,kpc from the centre of its cluster. It therefore seems unlikely
that we have missed many BCGs.

\subsection{Photometry}

Following \citet{Collins2009}, \citet{Stott2010}, and \citet{Lidman2012}, we use
{\tt MAG\_AUTO} in SExtractor to estimate the total Ks-band magnitude
of the brightest cluster galaxy in each cluster and use aperture
magnitudes to compute J-Ks colours. The colours are computed within a
16\,kpc diameter aperture after first matching the PSFs in the J and
Ks-band images.

The error in {\tt MAG\_AUTO} is estimated from the distribution of integrated
counts in circularised apertures that are randomly placed in regions
where there are no visible objects. The error in the aperture
magnitudes, which are used to derive colours, are estimated in the same
way. These errors are added in quadrature as an estimate of the error
in the colour. Further details on the methods used to compute the
magnitudes and colours of the BCGs can be found in
\citet{Lidman2012}. In Table~\ref{tab:BCGphot}, we list coordinates,
redshifts, colours and magnitudes of the BCGs. 

\begin{table*}
 \centering
 \begin{minipage}{140mm}
  \caption{J and Ks band photometry of the BCGs in the HCS clusters.\label{tab:BCGphot}}
  \begin{tabular}{lcclll}
  \hline
  Name  &   RA      &  Dec.  & Redshift & Ks      & J-Ks  \\
             & \multicolumn{2}{c}{[J2000]$^{\mathrm{a}}$}  & &  [mag] & [mag]\\

\hline
             RCS~231953+0038.0 & 23:19:53.43 & +00:38:13.4 & 0.9013 & 16.381 (0.007) & ... \\
             RCS~234526-3632.6 & 23:45:24.94 & -36:33:47.9 & 1.0380 & 17.411 (0.008) & ... \\
             RDCS~J1252.9-2927 & 12:52:54.42 & -29:27:17.6 & 1.2343 & 17.238 (0.017) & 1.888 (0.011) \\
               RX~J0152.7-1357 & 01:52:45.87 & -13:56:58.6 & 0.8342 & 16.657 (0.010) & ... \\
             XLSS~J0223.0-0436 & 02:23:03.72 & -04:36:18.2 & 1.2100 & 17.616 (0.008) & 1.932 (0.009) \\
             XMMU~J1229.4+0151 & 12:29:29.29 & +01:51:22.0 & 0.9740 & 17.255 (0.015) & ... \\
             XMMU~J2235.3-2557 & 22:35:20.85 & -25:57:39.8 & 1.3943 & 17.317 (0.026) & 1.960 (0.019) \\
           XMMXCS~J2215.9-1738 & 22:15:56.20 & -17:37:49.9 & 1.4545 & 18.650 (0.011) & 1.890 (0.016) \\
\hline
\end{tabular}
\begin{list}{}{}

\item[$^{\mathrm{a}}$] Coordinates are those of the BCG

\end{list}
\end{minipage}
\end{table*}

\begin{figure}
\includegraphics[width=9cm]{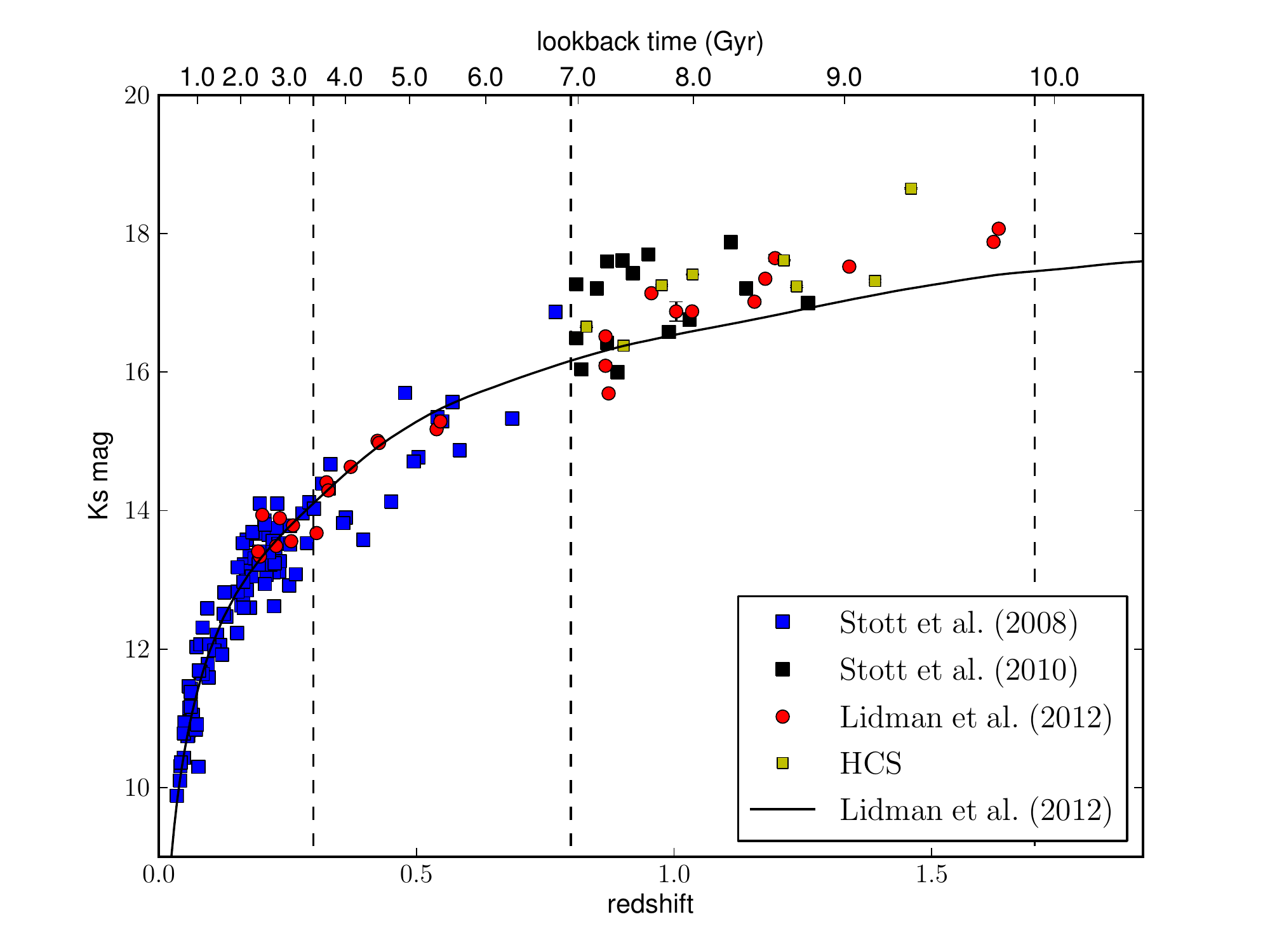}
\caption{The observer-frame Ks-band magnitude of BCGs as a function of
  redshift. The data from this paper are plotted as the yellow
  squares. Red circles beyond $z\sim 0.8$ are BCGs in the SpARCS
  clusters, while those below $z\sim 0.8$ are BCGs in the CNOC1
  clusters. The blue and black points are from \citet{Stott2008} and
  \citet{Stott2010}. The vertical dashed lines divide the sample into
  three redshift regions, labelled low, intermediate and high. The
  predicted Ks magnitudes of the best fit model used in
  \citet{Lidman2012} is shown as the black continuous
  line.  The normalisation of the model is
  set so that half of the data in the low-redshift bin lies above the model. Note how the data
  from this paper land within the region covered by data from earlier
  works and how most of the points in the high-redshift bin land above
  the model.}\label{fig:Hubble}
\end{figure}


\begin{figure}
\includegraphics[width=9cm]{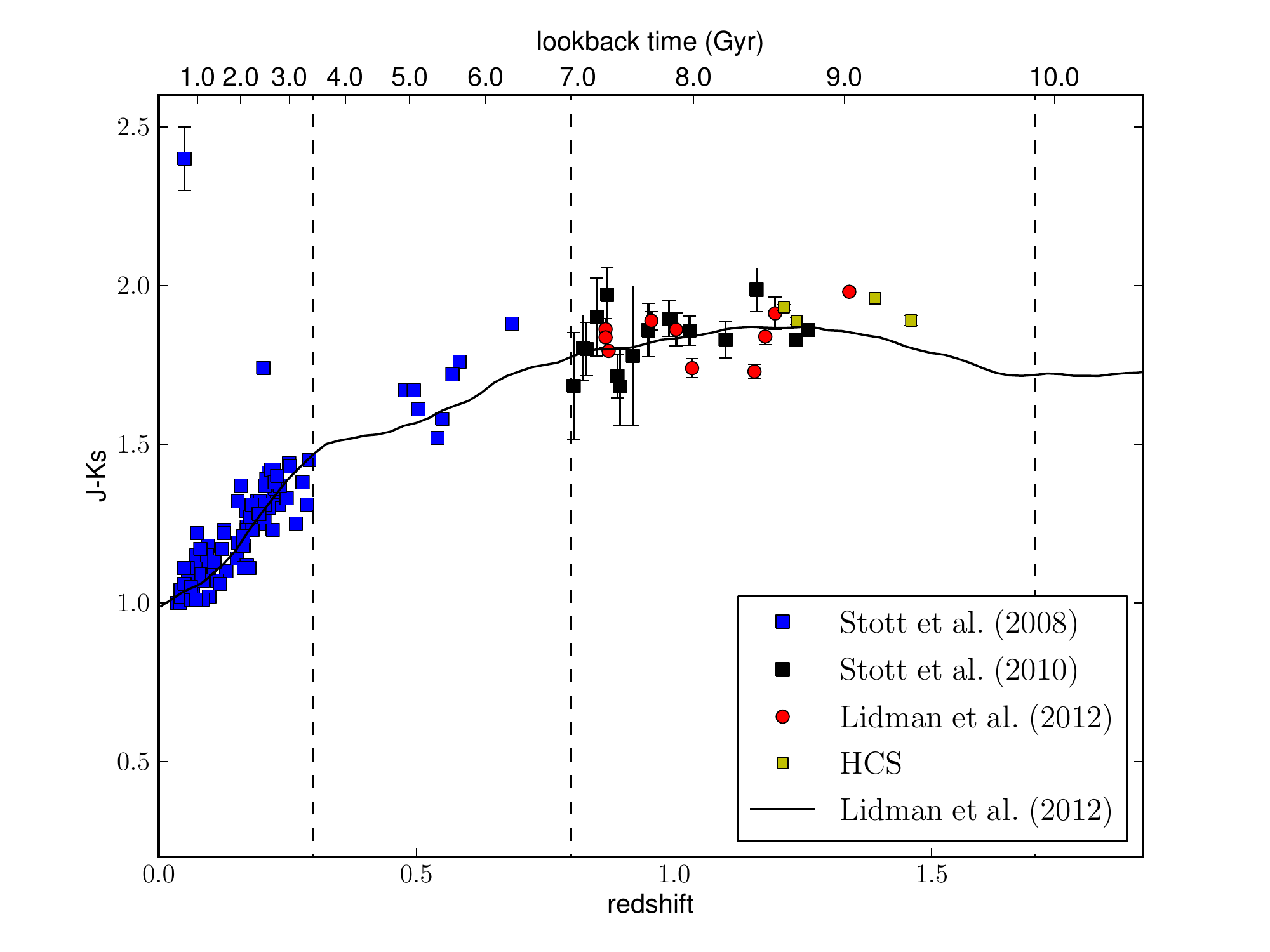}
\caption{The observer-frame J-Ks colour of BCGs as a function of
  redshift. The symbols have the same meaning as those in
  Fig.~\ref{fig:Hubble}. The blue point in the upper left-hand corner
  gives an indication of the uncertainty in the measurements from
  \citet{Stott2008}. Not all BCGs that were observed with HAWK-I have
  J band data from HAWK-I, so these BCGs are not shown, The black
  continuous line is the model that is used in \citet{Lidman2012} to
  match the observed colour over the entire redshift range covered by
  the data. The model adequately describes the average trend in colour
  from $z \sim 0$ to $z \sim 1.5$. See the text for a description of the model.}\label{fig:colour}
\end{figure}

In Fig.~\ref{fig:Hubble}, we plot the Ks-band magnitude of the BCGs in
our sample as a function of redshift. In this figure, we also plot the
Ks-band magnitude of BCGs from other clusters \citep{Collins2009,
  Stott2010, Lidman2012}. As noted in \citet{Lidman2012}, the BCGs in
the high-redshift subsample -- defined as BCGs with redshifts greater
than $z=0.8$ -- are systematically fainter than the model that best
describes the evolution in the J-Ks colour with redshift. The J-Ks
colours of the BCGs are shown in Fig.~\ref{fig:colour}. The model
is a composite of two models from \citet{Bruzual2003}, consisting
of a model that has  solar metallicity and a model that has a metallically that is
two--and--a--half times higher. Both models have an exponentially
falling star formation rate with an e--folding time of 0.9\,Gyr and a
formation redshift of $z=5$.  Additional details on the
  model can be found in \citet{Lidman2012}.

Five of the clusters in our sample were also observed in the Ks band
by \citet{Collins2009} and \citet{Stott2010} using MOIRCS on the
Subaru Telescope. \citet{Collins2009} and \citet{Stott2010} compute
magnitudes in exactly the same manner as we do here, so we can compare
the fluxes we derive with theirs.  We find a mean offset of 0.12 mag
with a dispersion of 0.04 mag, indicating a small
  difference between the two data sets. On average, the magnitudes
reported here are brighter.  Offsets in the image zero points is one possible
  explanation for the difference.  Differences in image quality between the two data sets
  is another, since image quality affects the size of the aperture
  that is used in computing {\tt MAG\_AUTO}.  The Ks-band image
  quality of the 5 clusters in common with the sample of clusters in
  \citet{Stott2010} ranges from 0\farcs31 to 0\farcs38, with a mean of
  0\farcs34. This compares to a mean image quality of $\sim$ 0\farcs5
  for clusters in \citet{Stott2010}. We degraded the image quality of
  our data to 0\farcs5 to see how {\tt MAG\_AUTO} for the BCGs changed. On
  average, the BCGs became 0.05\,mag, with considerable scatter between
  BCGs. The change exacerbates the disagreement in the photometry.

\citet{Lidman2012} used the difference between the model and the
observer-frame Ks-band magnitudes to estimate that the stellar mass of
BCGs increases by a factor of 1.8 between $z \sim 0.9$ and $z \sim
0.2$. Since all but four of the BCGs in this Fig.~\ref{fig:Hubble} are also in
\citet{Lidman2012} we do not repeat the analysis here.  We do note,
however, that the Ks-band magnitude of these BCGs (the BCGs in
RCS~2319, RCS~2345, and XMMU~1229 and RX~0152) are similar to the Ks-band
magnitude of the BCGs at $z\sim1$.

Instead, we examine the prevalence of close companions to the
BCG. \citet{Lidman2012} noted that merging might be the mechanism by
which BCGs accrued most of their stellar mass. By examining the
prevalence of close companions to BCGs in a large enough sample of
clusters, we might be able to see evidence of this.

\section{Bright nearby companions}\label{sec:companions}

\subsection{The observed number of bright nearby companions}

Direct inspection of the Ks-band images reveals that several BCGs
have nearby companions that are amongst the brightest galaxies in
their respective clusters. Examples include the BCG of RDCS~1252, in
which the 2nd brightest cluster galaxy is only 15\,kpc from the BCG,
and SpARCS~1616, in which the 3rd brightest cluster galaxy is 22\,kpc
from the BCG. 

\begin{figure}
\includegraphics[width=9cm]{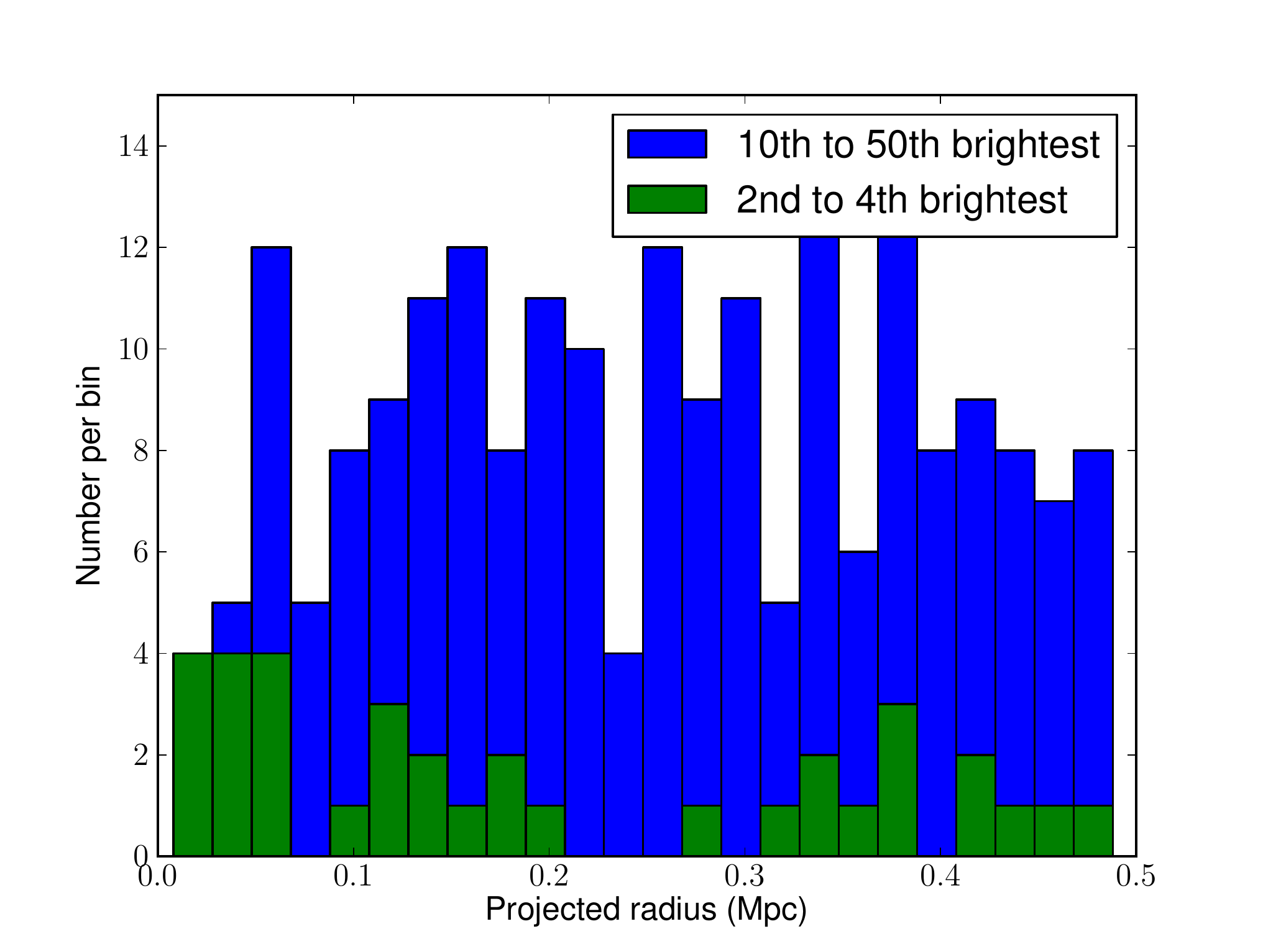}
\includegraphics[width=9cm]{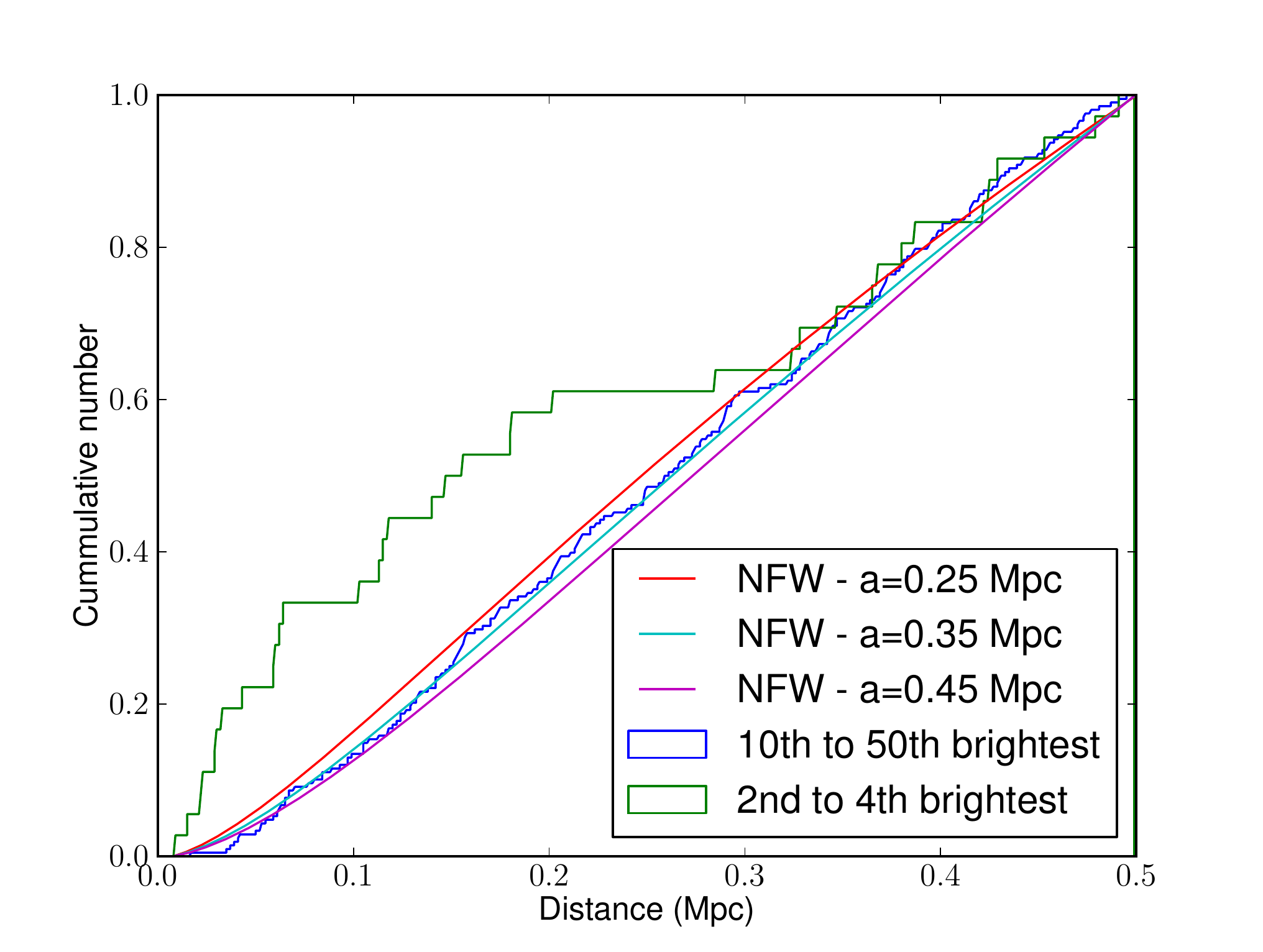}
\caption{{\bf Above:} Histograms showing the projected radial distance
  of the 2nd to 4th brightest cluster members (in green) and the
  10th to 50th brightest cluster members (in blue) from the
  BCG. The bin width is 20\,kpc {\bf Below}:
  The unbinned cumulative histogram of the two samples
  together with integrated NFW profiles with three different core radii. Note the
  difference between the two subsamples.}\label{fig:cumulative}
\end{figure}

To quantify this, we plot in Fig.~\ref{fig:cumulative} the number of
cluster galaxies -- all spectroscopically confirmed -- in annuli that
are 20\,kpc thick, starting from 8\,kpc from the BCG and extending out
to $\sim 0.5$\,Mpc. The lower limit in the first annulus corresponds
to the distance at which we can clearly separate galaxies down to the
magnitude limit that is probed by the spectroscopy. At the typical
redshift of the clusters, this corresponds to about 1\arcsec\ on the
sky. The plot is made for two subsamples: the 2nd, 3rd and 4th
brightest cluster members added together to make the first subsample
(green bins in Fig~\ref{fig:cumulative}), and all galaxies that are
between the 10th and 50th brightest cluster members (blue bins).

We have excluded any BCG that is more than 250\,kpc from the
luminosity weighted centre of its cluster. The excluded BCGs are the
BCGs of SpARCS~1051, SpARCS~1634, RCS~2345 and RX~0152. The distance
limit is chosen for a couple of reasons. Firstly, the limit has been
used in computing the merger rate at much lower redshifts
\citep{Edwards2012}, so we choose the same threshold to allow a more
direct comparison between the merger rates at low and high redshifts.
The fraction of excluded BCGs is $\sim 0.2$ and is similar to the
fraction excluded in \citet{Edwards2012}, but lower than the fraction
of non-central BCGs in \citet{Skibba2011}, who define non-central BCGs
differently. Secondly, it is possible that some of these far-flung
BCGs may not be the direct progenitors of the BCGs that we see at
lower redshifts, a point emphasised in \citet{DeLucia2007}. There is
observational evidence for this in at least one of the clusters. In
RX~0152, for example, it is likely that we would have identified a
more centrally located galaxy as the BCG if we had observed the
cluster a few 100\,Myr later.  The central region of the northern
clump in RX~0152 is dominated by two galaxies that are likely to merge
within a few 100\,Myr. If they did, the resulting galaxy might become
the BCG.  We will comment more on this interesting cluster later.

We note that by changing our selection so that the brightest galaxy in
the northern clump of RX~0152 is marked as the BCG, or by including
all BCGs outside the 250\,kpc distance limit does not alter our
conclusions. However, these BCGs are a potentially source of bias in
other studies. For example, the growth in the stellar mass of BCGs as
a function of redshift will be biased low if one includes galaxies
that do not become part of the BCG at a later time, even though these
galaxies are the brightest in the cluster at the time they were
observed.

It is apparent in Fig~\ref{fig:cumulative} that there is a significant
excess of galaxies that are classified as the 2nd, 3rd and 4th
brightest galaxies of the cluster in the inner three annuli (i.e.~out
to $\sim 70$\,kpc) compared to the annuli that are further out. The
excess is still visible if we consider just the 2nd brightest galaxy,
the 2nd and 3rd brightest galaxies, or the 2nd, 3rd, 4th and 5th
brightest galaxies added together. We will return to these bright
galaxies after examining the distribution of fainter galaxies.

In contrast to the brightest galaxies, the number of galaxies that are
fainter than the 10th brightest galaxy but brighter than the 50th is
much less concentrated. The inner most bin for this subsample has only
one object. Not all clusters have 50 spectroscopically confirmed
members. For these clusters we include all galaxies that are fainter than
the 10th brightest member.

A 2-sided KS test reveals that the probability (or P-value) that the
two samples are drawn from the same distribution is 1\%. This low
value provides support for rejecting the null hypothesis, i.e.~the
two samples are drawn from the same distribution. However, the
P-values are sensitive to the clusters used and to the radius out to
which the distributions are measured.  For example, limiting the
radius over which the histograms are compared to 300\,kpc results in a
P-value of 0.1\%.  On the other hand, restricting the faintest members
in the second sample to progressively brighter galaxies results in
higher P-values.  For these reasons, we take results of the KS test as
very suggestive rather than conclusive. The difference between the two
distributions may become clearer once a larger sample of clusters becomes
available.

The results of the KS test may be affected by differences in how the
spectroscopic completeness of the two subsamples changes with
projected radius.  The success of obtaining a spectroscopic redshift
depends on several factors, so the radial dependence of the
spectroscopic completeness may be different for bright and faint
objects.  To examine this issue, we compare curves that are obtained
by integrating projected Navarro-Frenk-White (NFW) profiles
\citep{Navarro1996,Bartelmann1996} with a range of core radii, $a$, to
the cumulative histograms of the two subsamples. The comparison is
made in the lower plot of Fig.~\ref{fig:cumulative}.  For the clusters in
our sample, these core radii correspond to concentrations ($c_{200} =
r_{200}/a$) ranging from 2.2 to 4, which covers the range expected
for massive clusters at $z\sim 1$ \citep{Duffy2008}. 

The radial distribution of the fainter cluster galaxies is consistent
with an NFW profile that has a core radius of $0.35$\,Mpc. A KS test
results in a P-value that is close to 1. There is no strong evidence
for a radial dependence in the spectroscopic
incompleteness in this subsample. The subsample containing the
brighter galaxies has a radial distribution that is considerably
different to that of the NFW profiles shown in
Fig.~\ref{fig:cumulative}. This could be interpreted as incompleteness
in the bright galaxy subsample at large radii. However, we consider
this explanation to be unlikely. Out to 250\, kpc, we
  estimate that the spectroscopic completeness for for objects that
  are brighter than 25\% of the Ks-band luminosity of the BCG is
  85\%. Out to 500\,kpc, this only drops to 78\%.


Instead, we believe that most of the difference between the two subsamples is
real and that the difference has a physical explanation. 
A plausible physical explanation for the difference between the bright
and faint samples is dynamical friction. Dynamical friction is more
effective in bringing large galaxies and galaxy groups into the core
region than it is in bringing in small galaxies. The radial
segregation between bright and faint clusters galaxies is observed in
nearby massive groups and clusters \citep{Pracy2005,Zandivarez2011}.
However, see \citet{Mei2007} for a different result in the Virgo cluster.





We now return our attention to the bright galaxies shown in
Fig.~\ref{fig:cumulative}.  The excess of galaxies in the innermost
annuli implies that the transverse velocities of most of these
galaxies are unlikely to be very high. If they were high, say
comparable to the velocity dispersion of the cluster, then the
excess in the inner most annuli would be erased. We note that the
circular speed of the NFW profile in this central region is about half
that at the core radius. If we make the additional assumption that
clusters are roughly spherical (an assumption that we will return to
later), then the excess also implies that most of these galaxies are
not strongly projected along the line--of-sight to the BCG.

The inner most annulus contains four galaxies.  The coordinates of
these galaxies, their Ks band flux relative to the BCG in their
cluster, their ranking in terms of brightness, and their line-of-sight
velocities with respect to their BCGs are shown in
Table~\ref{tab:companions}. Their proximity to the BCG means that
these galaxies could potentially merge with their respective BCGs in
less than 1\,Gyr \citep{Lotz2011}. There are no other galaxies that are within a factor
of four in mass with respect to the BCG (i.e.~a potential major merger)
and this close to the BCG.

\begin{table*}
\centering
\begin{minipage}{140mm}
  \caption{Bright cluster members that have a projected distance that is
    between 8 and 28\,kpc of the BCG. \label{tab:companions}}
 \begin{tabular}{lccllll}
 \hline
 Name  &   RA     &  Dec.  & Separation & Flux Ratio & Rank &Relative Velocity  \\
            & \multicolumn{2}{c}{[J2000]$^{\mathrm{a}}$}  &  [kpc] &
            & &  [km/s]$^{\mathrm{b}}$\\

\hline
             RCS~231953+0038.0 & 23 19 53.36 & +00 38 14.1 & 9.66 & 0.327 & 4 & 363 \\
             RDCS~J1252.9-2927 & 12 52 54.55 & -29 27 17.1 & 14.93 & 0.805 & 2 & 470 \\
             XMMU~J2235.3-2557 & 22 35 20.72 & -25 57 37.7 & 23.20 & 0.464 & 3 & 1415 \\
\hline
         SpARCS~J161641+554513 & 16 16 41.63 & +55 45 12.9 & 22.26 & 0.841 & 3 & 167 \\
\hline
\end{tabular}
\begin{list}{}{}
\item[$^{\mathrm{a}}$]Coordinates are those of the companion
\item[$^{\mathrm{b}}$]The relative velocity is the line--of--sight
  velocity difference between the companion and its BCG. For the HCS
  clusters, the uncertainty is around 100\,kms/s. For the SpARCS
  clusters, it is double this.

\end{list}
\end{minipage}
\end{table*}

\subsection{The expected number of bright nearby companions}

Given the timescale for how long it takes a bright nearby companion to
merge with the BCG, one can estimate the number of close companions one
should see if major mergers are the principle mechanism for the growth
in the stellar mass of BCGs. Since the timescales are only approximately
known, we examine a range of timescales.  At the lower end is the time
it takes for two galaxies to merge once they are already in the
process of merging. This is of the order of a few crossing times
\citep{Binney2008}.  At a distance of 30\,kpc, the crossing time for a
BCG with a velocity dispersion of 250 km/s is around 70\,Myr,
resulting in a merging timescale of around 200\,Myr.  It is over this
sort of timescale that one would expect to see direct evidence of the
merger occurring, i.e. evidence of diffuse tidal tails, highly
distorted isophotes and broad fans.


When direct evidence for merging is not apparent, the timescale will be larger.
\citet{Lotz2011} derive a merger timescale of 600\,Myr for pairs of
field galaxies that have a projected separation that ranges between 7
and 21\,kpc, a velocity difference that is less than 500 km/s and a
mass ratio that is in the major merger mass range. 

\citet{Kitzbichler2008} provide a formula to compute the timescale
for merging for galaxies that have a line--of--sight velocity
difference that is less than of 300\,km/s. For a pair of galaxies at
$z\sim1$ with a projected distance of 21\,kpc and a combined stellar mass 
of $5\times10^{11}\,M_{\odot}$, the timescale is 500\,Myr, which is
similar to the timescale found by \citet{Lotz2011}.

The timescales derived by \citet{Lotz2011} and \citet{Kitzbichler2008}
apply equally well to galaxy pairs in the centre of clusters (we return
to this point in Sec.~\ref{sec:offsets}). By Newton's
first theorem, mass outside the orbit of the galaxy pair is not felt.
For our work, the timescales of \citet{Lotz2011} and \citet{Kitzbichler2008} are the
most appropriate ones to use. However, we also examine what happens if the
timescale is considerably longer, i.e.~1.0\,Gyr.

\begin{figure}
\includegraphics[width=9cm]{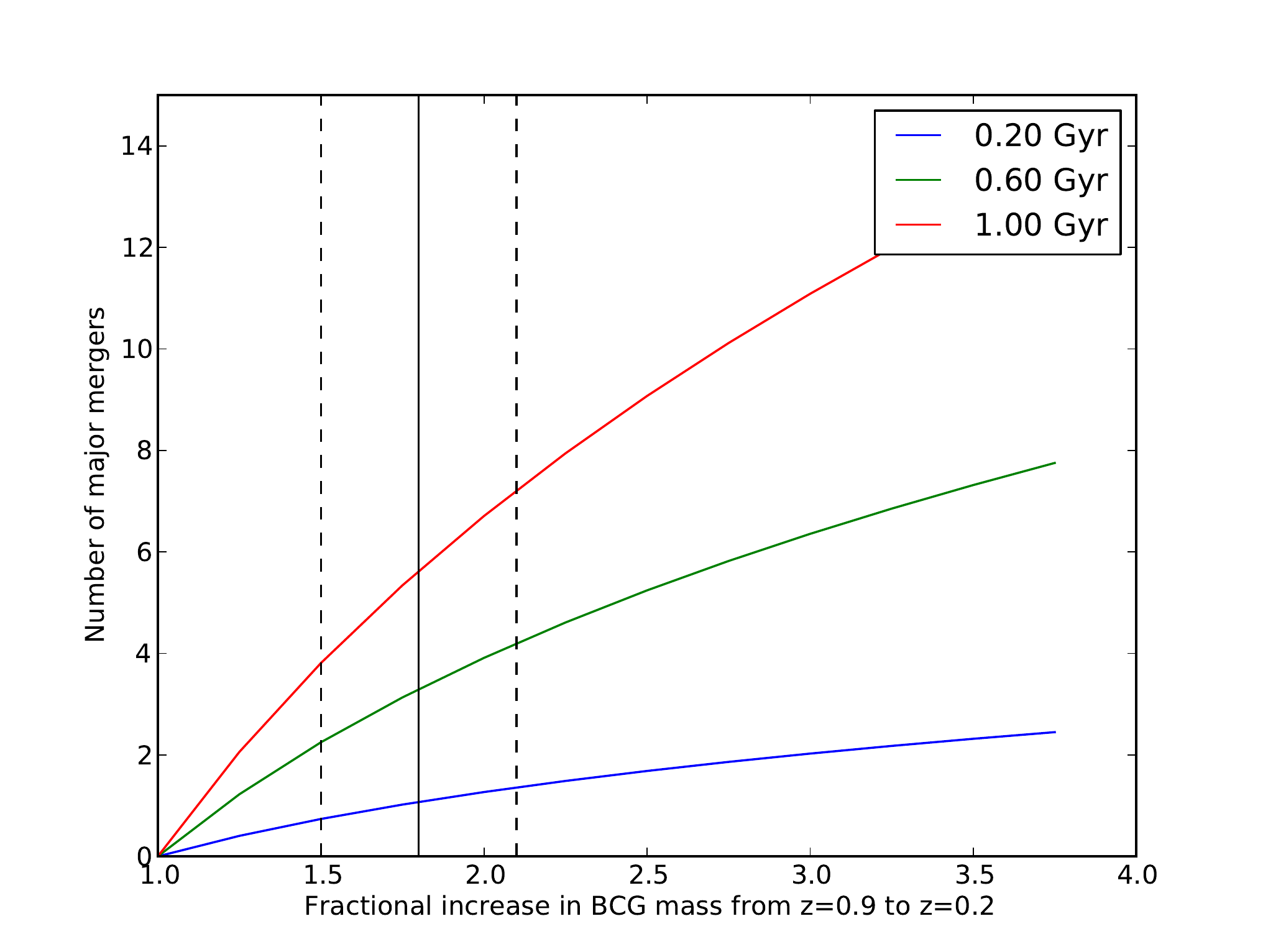}
\caption{The number of major mergers one would expect to see
  in our sample of 14 high redshifts clusters as a function of how
  much the BCG grows between redshift $z=0.9$ and $z=0.2$. The
  number is plotted for three merger timescales: 0.2\,Gyr,
  0.6\,Gyr and 1.0\,Gyr. The vertical solid line represents the growth
  that has been measured by observations and the
  dashed line is the uncertainty \citep{Lidman2012}. For a timescale
  of 0.6\,Gyr, we would expect that 3 major mergers would occur in our 14 high redshift
  clusters within 600\,Myr. This is remarkably close to what we observe.}\label{fig:expected}
\end{figure}

In Fig.~\ref{fig:expected}, we plot the number of major mergers one
would expect to see in our sample of 14 high redshifts clusters as a
function of how much the BCG grows between redshift $z=0.9$ and
$z=0.2$. We make the assumption that the major merger rate is constant with redshift. We compute the number for three timescales, 200\,Myr,
600\,Myr and 1\,Gyr. The vertical line represents the growth that
has been measured by observations \citep{Lidman2012}.  In this plot,
we have assumed that all the mass growth is due to mergers with
galaxies that are 62.5\% of the mass of the BCG\footnote{This is
  simply the average mass of the companion in a 1:1 merger and  the
  companion in a 1:4 merger}, and that
50\% of the companion is accreted onto the BCG. High resolution
simulations suggest that between 50 to 80\% of the mass of mergers
will be distributed throughout the cluster
\citep{Conroy2007,Puchwein2010} and visible as intra-cluster
light. If only 20\% of the mass of the companion is accreted onto the
BCG, these curves move up by a factor of 2.5.

The vertical line intersects the middle green curve in
Fig~\ref{fig:expected}), which represents a merging timescale of
600\,Myr, at $\sim 3$. In other words, one would expect to see
evidence for 3 major mergers in the 14 distant clusters of our sample.
Interestingly, there are three galaxies in Table~\ref{tab:companions}
that we identify as galaxies that will merge with the BCG within
600\,Myr: one of each in RDCS~1252, SpARCS~1616, and RCS~2319. The galaxy
in XMMU~2235 is moving too fast with respect to the BCG to merge with
the BCG over this timescale.

We can use Fig.~\ref{fig:expected} to estimate the number of major
mergers that are expected to occur within the next 200\,Myr. This is
the timescale over which one should see evidence of an interaction
through, for example, diffuse fans, distorted isophotes and tidal
tails. Using Fig.~\ref{fig:expected}, we find that we should see one
such case in our sample of 14 clusters. Interestingly, there is one
pair  where there is evidence of an ongoing interaction. This is
the pair in RDCS~1252 \citep{Blakeslee2003,Rettura2006}. 





\section{Discussion}\label{sec:discussion}

\subsection{Sources of Uncertainty}\label{sec:uncertainty}

There is, of course, considerable uncertainty in the estimated number
of major mergers. First and foremost, there is the uncertainty that
comes from small number statistics. Out of 14 clusters that were
examined, we find that there are three galaxies that are likely to
result in a major merger with the BCG within 600\,Myr. We use the beta
distribution to compute confidence intervals \citep{Cameron2011} on
the probability that a cluster contains a galaxy pair of this type at
any one time. The 68.3\% confidence interval goes from 0.14 to 0.36.
This is a very broad range.



There is also the possibility that some of the galaxies that we have
assumed to be near to the BCG are in fact projections along the
line--of--sight. The probability of this increases if clusters are
extended along the line--of--sight. All of our clusters are selected
either as over-densities in red sequence galaxies (e.g. the clusters
from SpARCS and RCS) or as extended X-ray sources
(e.g. RDCS~1252). Both types of selection preferentially select
clusters that are extended along the line--of--sight. 

The cluster RDCS~1252 is an interesting case. By analysing the angular
structure of cluster members together with their velocities,
\citet{Demarco2007} conclude that RDCS~1252 consists of two sub-clusters
in the process of merging, with the BCG of the cluster centered in one
of the sub-clusters and the 2nd brightest galaxy centered in the
other. RDCS~1252 provides support for the idea that some clusters are
extended along the line--of--sight. In such cases, the two brightest galaxies 
in the cluster may be much further apart than we expected.

Nevertheless, the central pair of galaxies in RDCS~1252 provide us with
more information.  \citet{Blakeslee2003} and \citet{Rettura2006} both
find evidence of an S-shaped residual linking the centre of the
two brightest galaxies in the model-subtracted images, thus providing a clear sign
that the galaxies are merging and are indeed near to each other.

\subsection{Velocity offsets}\label{sec:offsets}

In computing the merger timescale for the BCG and its companion, we
have argued that we can use the timescales that have been computed for
field galaxies. This is only valid if the BCG and its companion are centered in the
cluster and at rest with respect to it. If these conditions are not
met, then the BCG and its companion would be subject to tidal forces from the cluster. This
would increase the merger timescale and perhaps even prevent the
BCG and its companion from merging \citep{Mihos2003}.

We examine the spatial and dynamical properties of the three BCGs --
the BCGs in SpARCS~16116, RDCS~1252 and RCS~2319 -- that we have
identified as undergoing a potential major merger within $\sim
500$\,Myr. All three are spatially well centered in their respective
clusters. Dynamically speaking, we can only compare the redshift of
the BCG with respect to median redshift of the cluster galaxies, which
we will use as a proxy for the cluster redshift. As can be seen in
Table \ref{tab:BCGvel}, within errors, the velocity offset of the BCGs
in SpARCS~1616 and RCS~2319, are consistent with zero, meaning that
the BCGs are practically at rest with respect to the cluster. Using
the merger timescale that has been used for field studies is therefore reasonable.

The BCG of RDCS~1252, however, is not at rest with respect to its
cluster. As noted in the previous section, RDCS~1252 appears to
consist of two merging sub-clusters \citet{Demarco2007}, with the BCG
centered in one sub-cluster and its companion
centered in the other. It seems likely that cluster tides will play an
important role in setting the timescale over which these two galaxies
will merge. There seems little doubt that they will merge, as there is
evidence of an interaction between the BCG and the 2nd brightest
galaxy in this cluster \citep{Blakeslee2003,Rettura2006}

Looking at the entire sample, we see that a few other BCGs are not at rest within
their respective clusters. The most notable of these is the BCG in
SpARCS~0035. Such offsets are indicative of a cluster merger. SpARCS~0035
is discussed further in Rettura et al. (in preparation).

\begin{table}
 \centering
 \begin{minipage}{140mm}
  \caption{Velocity offset between the cluster redshift and the BCG\label{tab:BCGvel}}
  \begin{tabular}{lr}
  \hline
  Short Name  &  Velocity Offset \\
& [km\,s$^{-1}$] \\
\hline
               RX~J0152.7-1357 &   290 $\pm$   240\\
             RCS~231953+0038.0 &   170 $\pm$   190\\
             XMMU~J1229.4+0151 &   230 $\pm$   210\\
             RCS~234526-3632.6 &  -290 $\pm$   160\\
             XLSS~J0223.0-0436 &   430 $\pm$   200\\
             RDCS~J1252.9-2927 &   500 $\pm$   160\\
             XMMU~J2235.3-2557 &  -540 $\pm$   210\\
           XMMXCS~J2215.9-1738 &   670 $\pm$   170\\
\hline
         SpARCS~J003442-430752 &  -140 $\pm$   220\\
         SpARCS~J003645-441050 &   300 $\pm$   230\\
         SpARCS~J161314+564930 &  -180 $\pm$   230\\
         SpARCS~J104737+574137 &    80 $\pm$   220\\
         SpARCS~J021524-034331 &   -60 $\pm$   180\\
         SpARCS~J105111+581803 &   -40 $\pm$   190\\
         SpARCS~J161641+554513 &   -30 $\pm$   190\\
         SpARCS~J163435+402151 &    30 $\pm$   190\\
         SpARCS~J163852+403843 &  -100 $\pm$   190\\
         SpARCS~J003550-431224 &  1030 $\pm$   240\\
\hline

\end{tabular}
\end{minipage}
\end{table}

\subsection{RX~0152}

In Sec~\ref{sec:companions}, we excluded four clusters because the BCG was
more than 250\,kpc from the luminosity--weighted centres of the clusters.
One of the excluded clusters was RX~0152. 

\begin{figure*}
\setlength{\unitlength}{0.3125mm}
\begin{picture}(512,256)
\includegraphics[width=16cm]{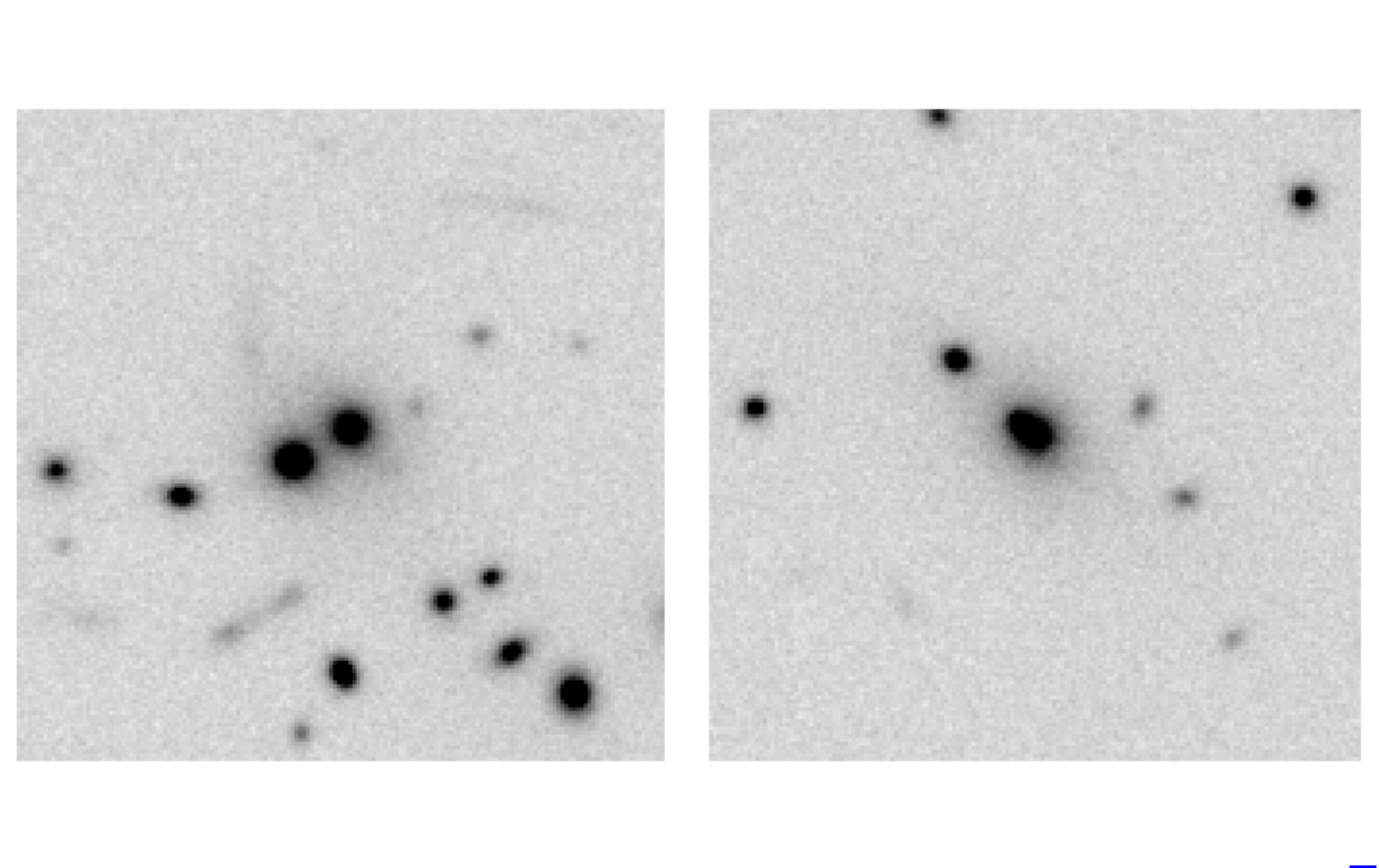}
\put(-495,50){\Large{Northern clump}}
\put(-245,50){\Large{Southern clump}}
\end{picture}
\caption{HAWK-I Ks-band images of the central galaxies in the two
  largest clumps of RX~0152 \citep{Demarco2005}. {\bf Left} the
  central region of the northern clump. Note the faint, gently curving
  material towards the north-east of the pair.  In the higher
  resolution ACS images, it seems to be a smaller galaxy that is
  interacting with the two brighter galaxies. Alternatively, it could
  be a strongly lensed background galaxy.  {\bf Right} the central
  region of the southern clump. Note the very close companion to the
  central galaxy of the southern clump, which can be seen as an
  extension of the isophotes to the north-east of the galaxy. Nantais
  et al. (in preparation) have noted that this object is very compact
  in the ACS images, which is an indication that it is an advanced
  state of merging with the bigger galaxy. The images are 18\arcsec\
  on a side, which corresponds to 140\,kpc at the redshift of the
  cluster. North is up and East is to the left.}\label{fig:rx0152}
\end{figure*}

As noted earlier, RX~0152 consists of at least three dynamically
distinct clumps \citep{Demarco2005, Demarco2010b}. The BCG is not
centered in any of the clumps: instead, it is centered about
35\arcsec\ (270\,kpc) north-east of the northern clump. Closer
inspection of the galaxies that are centred in the two biggest clumps
reveals that both have bright, nearby companions, as illustrated in
Fig.~\ref{fig:rx0152}.

Both galaxies in the northern clump are cluster members. They are
separated by 12\,kpc and differ in velocity by 280\,km/s. They almost
have the same brightness, differing in Ks band flux by only 1\%. Using
the arguments presented in Sec.~\ref{sec:companions}, it is likely that
these galaxies will merge within a few 100\,Myr. If these galaxies do
merge and if all of the mass stays in the descendent of the merger,
then the descendent will become the brightest galaxy in the cluster.
Intriguingly, in the deep HAWK-I data, there appears to be material
possibly extending from this pair to the north-east. The material is
also visible in the ACS image. It may be a background arc, although
the orientation is more radial than tangential, or it could be
material from a third galaxy that is in the process of being destroyed
by its two much more massive companions.

The central region in the southern clump is actually a triplet of
galaxies, consisting of a pair of cluster members that are 20\,kpc
apart and a 3rd galaxy that is only 3.5\,kpc from the brightest galaxy
in the triplet, which was the galaxy chosen as the BCG in
\citet{Stott2010}. Unfortunately, we do not have a redshift for the
3rd galaxy. However, as noted in Nantais et al. (in preparation), the
third galaxy in the ACS images appears to be very compact
when compared to cluster galaxies of similar brightness. This is
indicative of an ongoing merger.  The halo of the galaxy has been
stripped by its brighter neighbour. The redshifts of the other two
galaxies are, within the errors, identical, which means that the relative
velocity of the galaxies differ by less than 100\,km/s. These two
galaxies could also merge within a few 100\,Myr. If all three galaxies
were to merge and if there was no subsequent star-formation, then the
brightness of the resulting galaxy would be comparable to the
brightness of what is now the BCG.

\subsection{Comparison with low-redshift subsamples}\label{sec:low-z}

A number of studies have estimated the number of bright
galaxies that are close to BCGs in nearby galaxy clusters. \citet{Liu2009}
estimated that about 49 BCGs out of a sample of 515 BCGs in the
redshift interval $0.03 \le z \le 0.12$ have a
companion that is i) within a projected distance of 7 to 30\, kpc of
the BCG and ii) within 2 mag of the SDSS r-band magnitude of the BCG. They
apply a further restriction, which applies to both the BCG and its
companion. The g-r colour of both galaxies must be greater than 0.7.

The criteria adopted by \citet{Liu2009} are broadly similar to the ones used in
this paper. They find that 1 BCG in 10 have a bright nearby companion,
whereas we find that 3 BCGs in 14 have a nearby bright companion. The
number of companions in our more distant clusters is higher; however,
given the small number of clusters in the high redshift sample, it is
premature to cite this as strong evidence that the number of bright
nearby companions to BCGs in distant clusters is increasing with
redshift. We return to this point in the following section.

\citet{Liu2009} then go on to examine the number of BCGs that show
direct evidence of merging with their nearby companion. Evidence of a
merger includes tidal tails, distorted isophotes, and broad fans. Out of 49
close pairs, 18 or one-third show evidence for a merger taking
place. While we again stress that the number of BCGs in our
high-redshift sample is small, one of our three pairs, the pair in
RDCS~1252, shows evidence for a merger.

\citet{Liu2009} estimate that BCGs are increasing their stellar mass
at a rate of 2.5\% $(t_{\mathrm{merger}} / 0.3 \mathrm{Gyr})^{-1}
(f_{\mathrm{mass}}/0.5)$, where $t_{\mathrm{merger}}$ is the timescale
for the merger to take place and $f_{\mathrm{mass}}$ is the fraction
of companion's stellar mass that is accreted onto the BCG.  In a
separate study, \citet{Edwards2012} find that BCGs at $z \sim 0.3$ are
adding as much as 10\% of their stellar mass through mergers, both
minor and major, over 0.5\,Gyr, which is considerably higher than the
rate in \citet{Liu2009}.  However, recall that \citet{Liu2009} only
include cases where there is direct evidence of a merger and only
major mergers. The two results differ because of the broader
definition of potential mergers and the broader mass range used in
\citet{Edwards2012}. They also probe different redshifts.

In our study of 14 high redshift clusters, we find that BCGs are
accreting mass at a rate of 7\% $(t_{\mathrm{merger}} / 0.6
\mathrm{Gyr})^{-1}$. This estimate assumes that half of the mass of
the companion is accreted onto the BCG and half is distributed more
broadly throughout the cluster. Our rate is higher than that
inferred by \citet{Liu2009}.  On the other hand, it is lower
than the rate computed for low redshift clusters in
\citet{Edwards2012}; however, we do not include minor mergers, whereas
\citet{Edwards2012} do, so part of the difference is in part due to
the broader mass range used in \citet{Edwards2012}.

While our sample is too small to test for a change in the major merger
rate with redshift, it is clear that major mergers are occurring at $z
\sim 1$ and that the rate is comparable to the rate at lower redshifts. If major mergers contribute as much
mass as minor mergers, and N-body simulations suggest that they
contribute more \citep{Laporte2013,Hopkins2010a}, then major mergers are not only
an important mechanism for the build up of stellar at $z\sim1$, they
are an important mechanism between $z\sim1$ and today.

Theoretically, one would expect major mergers to be more common at
higher redshifts. The accretion rate for massive clusters peaks
between $z=1.5$ and $z=2$ \citep{Fakhouri2010}, and as clusters get
larger, the dynamical friction timescale for galaxies of a given mass
increases, so it takes longer for a galaxy to sink into the core. For
these reasons, one would expect BCGs to experience fewer major mergers
at lower redshifts. We now compare our results with the results from
models.

\subsection{Comparison with models}

\citet{Hopkins2010a} use semi-empirical models to estimate the major
merger\footnote{Note that \citet{Hopkins2010a} use a more
restrictive range of mass ratios than we do here. They define a major
merger as one in which $\mu_{\star}$ lies within the range $0.33
<\mu_{\star} <1$.} rate as a function of stellar mass and redshift. At $z\sim 1$,
they find that the major merger rate per galaxy varies between $\sim 0.15$
to $\sim 0.4$\,Gyr$^{-1}$ for galaxies with stellar masses between
$10^{11 }M_{\odot}$ and $10^{12} M_{\odot}$.  The merger rate increases quickly
with stellar mass and with redshift. 

This mass range covers the range of stellar masses of the BCGs in our
sample, which varies between $10^{11} M_{\odot}$ and $8 \times 10^{11}
M_{\odot}$, with a median mass of $\sim 3 \times 10^{11} M_{\odot}$. Our BCGs
vary in redshift from $z=0.84$ to $z=1.46$, with a median of
$z\sim1.1$.

If we use the more restrictive definition of what constitutes a major
merger used in \citet{Hopkins2010a}, we find that 2 out of 14 of the
BCGs in our sample -- the 3rd galaxy has a mass ratio 0.327, and
therefore just fails to meet the more restrictive definition
-- are likely to experience a major merger within 600\,Myr. This
translates to a rate of 0.25 major mergers per Gyr, which is fully
consistent with the rates derived in \citet{Hopkins2010a}. Using the
broader definition that we have used throughout the paper, we find a rate
of 0.4 major mergers per Gyr. 

As noted in Sec.~\ref{sec:low-z}, we find a higher fraction of bright
nearby companions in our sample than others have found in samples at lower
redshifts. We find them to be about a factor of about two more common. 
Over the stellar mass range covered by the BCGs in our sample,
\citet{Hopkins2010a} find that the major merger rate increases between
$z=0$ and $z=1$ by a factor that varies between 2 and 4. This is
consistent to what we infer from observations.

While there is consistency between models and observations, one needs
to be mindful of potential biases that come from the way clusters are
selected at low and high redshifts. The clusters in our sample are
some of the most massive clusters that we know of at these high
redshifts. By today, they would have increased in mass significantly,
from a median mass of $3.6 \times 10^{14}\mathrm{M}_{\odot}$ at $\sim1.1$ to a
median mass of $1.3 \times10^{15}\mathrm{M}_{\odot}$. By today, the average
cluster in our sample of high redshift clusters would be more massive
than the average cluster in the sample used in, for example, \citet{Liu2009}. Since
the properties of the BCG correlates with the properties of the
cluster in which it lives (e.g.~more massive clusters have more
massive BCGs), it is not unreasonable to expect that the major merger
rate does too. Comparing this rate in cluster samples that span
different mass ranges may lead to a biased view of how this rate changes
with redshift. Currently our samples are too small to explore how
significant this bias may be. 

\section{Summary and Future Work}

We combine 10 distant clusters from SpARCS with the 9 clusters of the
HAWK-I cluster survey to build a sample of 19 galaxy clusters between
$z=0.84$ and $z=1.46$. Our sample contains over 600 spectroscopically
confirmed cluster members. We use this sample to examine the frequency
of bright cluster members that are likely to merge with the BCG within
600\,Myr.

After excluding one cluster because of the uncertainty in identifying
the BCG -- due to the chance projection of a much nearer face-on
spiral galaxy close to the cluster core -- and four others because the
BCGs are located more than 250\,kpc from the cluster centres, we find that 3
of the 14 BCGs are likely to experience a major merger within
600\,Myr. While the statistical uncertainty stemming from the small
number of clusters in our sample is large, the number of mergers is
similar to the number of mergers that are predicted by theory and to
the number that would be needed to build the stellar mass of BCGs by a
factor of $\sim 2$ between redshift $z=0.9$ and $z=0.2$, under the
assumptions that major mergers contribute most of the accreted stellar
mass and that half of the mass of the companion is accreted onto the
BCG. The factor of two increase in the stellar mass between redshift
$z=0.9$ and $z=0.2$ has been measured from observations
\citep{Lidman2012} and predicted by N-body models \citep{Laporte2013}.

The data are consistent with the notion that the majority of the
stellar mass that is accreted onto BCGs between $z \sim 1$ and today
comes from major mergers. However, they do not exclude the possibility
that minor mergers could play an important role in shaping how BCGs
appear today. In future work (Rettura et al. in preparation), we will
examine the size of the BCGs in our sample and compare them to BCGs at
lower redshifts. It is expected that major mergers increase the size
of the galaxy linearly with the amount of mass accreted, whereas minor
mergers are expected to increase the size of the galaxy more quickly
than this. 

\section*{Acknowledgments}

The authors thank Chris Collins, John Stott, Adam Duffy and Darren Croton for useful discussions,
and Adam Stanford for making available the redshifts for XMMXCS~2215.
The data in this paper were based in part on observations obtained at
the ESO Paranal Observatory (ESO programmes 060.A-9284(H) and 084.A-0214). C.L. is the
recipient of an Australian Research Council Future Fellowship (program
number FT0992259).  L.F.B. was supported by FONDECYT grants No. 1085286
and 1120676. W.J.C. is the recipient of an Australian Research
Council Professorial Fellowship (program number
DP0877642). J.N. acknowledges the support provided from FONDECYT
research grant number 3120233. R.D. acknowledges the support provided
from FONDECYT research grant number 1100540. G.W. gratefully
acknowledges support from NSF grant AST-0909198.
 


\label{lastpage}

\end{document}